\documentclass[11pt]{article}
\usepackage{graphicx}

\begin{document}

\begin{center}
{\bf \large Physical parameters of close binary systems: VI}
\end{center}

\begin{center}
by
\end{center}

\begin{center}
\large K.D.~Gazeas$^{1}$, P.G.~Niarchos$^{1}$, S.~Zola$^{2,3}$, J.M.~Kreiner$^{3}$, S.M.~Rucinski$^{4}$
\end{center}

\begin{center}

$^1$  Department of Astrophysics, Astronomy and Mechanics, University of Athens, Panepistimiopolis, 
      GR-15784 Zografos,  Athens, Greece\\ 
      e-mail: pniarcho@cc.uoa.gr \\

$^2$ Astronomical Observatory of the Jagiellonian University, ul. Orla 171, 30-244 Cracow, Poland \\
     email: sfzola@cyf-kr.edu.pl \\
 
$^3$ Mt. Suhora Observatory of the Pedagogical University, ul. Podchor\c{a}\.{z}ych 2, 30-084 Cracow, Poland \\
     email: sfkreine@cyf-kr.edu.pl\\
    
$^4$ David Dunlap Observatory, University of Toronto, P.O. Box 360, Richmond Hill, Ontario, Canada L4C 4Y6 \\
     email: rucinski@astro.utoronto.ca \\

\end{center} 
 
\begin{center}
ABSTRACT
\end{center}

New high-quality CCD photometric light curves for the W UMa-type systems V410~Aur, CK~Boo, 
FP~Boo, V921~Her, ET~Leo, XZ~Leo, V839~Oph, V2357~Oph, AQ~Psc and VY~Sex are presented. 
The new multicolor light curves, combined with the spectroscopic data recently obtained 
at David Dunlap Observatory, are analyzed with the Wilson-Devinney code to yield the 
physical parameters (masses, radii and luminosities) of the components.
Our models for all ten systems resulted in a contact configuration. Four binaries  
(V921~Her, XZ~Leo, V2357~Oph and VY~Sex) have low, while two (V410~Aur and CK~Boo) have 
high fillout factors. FP~Boo, ET~Leo, V839~Oph  and AQ~Psc have medium values of the 
fillout factor.
Three of the systems (FP~Boo, V921~Her and  XZ~Leo) have very bright primaries as a result  
of their high temperatures and large radii.\\

{\noindent}

\begin{center}
{\bf 1. ~Introduction}
\end{center}

In this paper we present the combined multicolor photometric and spectroscopic study 
of ten contact systems from the sample defined by Kreiner et al. (2003).
The systems selected to be studied are: V410~Aur, CK~Boo, FP~Boo, V921~Her, ET~Leo, 
XZ~Leo, V839~Oph, V2357~Oph, AQ~Psc and VY~Sex. 
The targets were selected according to the accurate multicolor photometric light curves, 
obtained recently from ground-based observations, and the spectroscopic mass ratio, 
derived from the radial velocity curves from the DDO program (Rucinski et al. 2005 and 
references therein).

Details of the project and the procedure used to derive the parameters were given in  
Kreiner et al. 2003 (Paper~I) and in Baran et al. 2004 (Paper~II).
The selected systems are analysed with the same methods and techniques, as those in 
Zola et al. 2004 (Paper III),  Gazeas et al. 2005a (Paper~IV) and Zola et al. 2005 (Paper~V).
Remarks on individual systems are given in Section 2, while Section 3 contains the log 
of the new photometric observations and description of the instruments used. Section 4 
outlines the procedure and method used to analyze the photometric and spectroscopic data, 
and to compute the absolute parameters of the components. 
A discussion of the results for each system is given in the last section. The theoretical 
light curves are compared with the observed ones and the assumptions 
made to explain better the observations, such as the existence of a third body in a system 
or magnetic activity of the components, are also discussed.\\

\begin{center}
{\bf 2. ~Remarks on the individual systems}
\end{center}

{\it 2.1. ~V410~Aur} \\   
The eclipsing binary V410~Aur (BD $+$34$^{\circ}$ 934, HIP 23337, V = 10.33$^{m}$) was 
discovered by Hipparcos satellite mission (ESA, 1997). It has a typical light curve of 
a W UMa-type  binary with total eclipses at the secondary minimum. There are limited 
observations of this system, such as those obtained by Dvorak (2005), who gave a time of 
minimum. 
V410~Aur is a spectroscopicaly triple system with a third component having relative brightness 
of  L$_{3}$/(L$_{1}$+L$_{2}$)=0.26 $\pm$ 0.01. Rucinski et al. (2003) 
calculated the mass ratio $q_{sp}$=0.144 $\pm$ 0.013 and classified the system as 
an A-subtype G0/2V contact binary. Yang et al. (2005) observed this binary and 
calculated its absolute photometric parameters. They also determined the mass ratio 
photometrically, giving the value of $q_{ph}$=0.1428 and found a filling factor 
of $52.4\%$.
Finally, they suggested that the system undergoes mass transfer from the less massive 
component to the more massive one.\\

{\it 2.2. ~CK~Boo} \\
The eclipsing binary CK~Boo (BD $+$9$^{\circ}$ 2916, HIP 71319, V = 9.09$^{m}$) was 
discovered by Bond (1975). The light variability was determined to be typical of a 
W~UMa-type binary. Krzesinski et al. (1991) gave a solution with low inclination and partial 
eclipses. Rucinski and Lu (1999) calculated the mass ratio of this A-subtype contact binary 
$q_{sp}$ = 0.111 $\pm$ 0.052 and the spectral  type F7/8V. Aslan and Derman (1986) presented
an extended research of CK~Boo based on observations obtained between 1976 and 1982. 
They noticed intrinsic light curve variations, mostly seen in the primary minimum, 
as well as variations of the color of the system. 
Qian and Liu (2000) published a list of minina of CK~Boo and discussed a possible connection 
between the variation of light curve and the change of the orbital period. 
The most recent study of CK Boo was published by Kalci and Derman (2005). 
They performed light curve modelling using new BVR light curves. Due to well
pronounced O'Connell effect a model with a cool spot on the surface of the
larger component was adopted. Combined their results with spectroscopic data by Rucinski
and Lu (1999) the authors derived the absolute parameters of the components.\\

{\it 2.3. ~FP~Boo} \\
FP~Boo (BD $+$43$^{\circ}$ 2523, HIP 76970, V = 10.20$^{m}$) was discovered by the Hipparcos 
satellite mission. These  data were analyzed by Selam (2004) who obtained the photometric 
mass ratio of the system to be $q_{ph}=0.1$, which is in good agreement with the spectroscopic 
one ($q_{sp} = 0.106 \pm 0.005$) derived by Rucinski et al. (2005). The small mass ratio 
of the system and the large radial velocities may indicate a high inclination system 
with total eclipses, similar to the contact binary AW~UMa.\\

{\it 2.4. ~V921~Her} \\
The light variability of the eclipsing binary V921~Her (BD $+$47$^{\circ}$ 2388, HIP 82344, 
V = 9.49$^{m}$) was also discovered by the Hipparcos mission. It has a relatively 
long period (P=0.877 days), rare for a contact binary. V921~Her has a low amplitude of light 
variation and small RV values, indicating a low inclination orbit. Rucinski et al. (2003) 
determined the mass ratio $q_{sp}$ = 0.226 $\pm$ 0.005 and classified the system as an 
A-subtype A7IV contact binary.\\

{\it 2.5. ~ET~Leo} \\
ET~Leo (BD $+$18$^{\circ}$ 2374, HIP 51677, V = 9.60$^{m}$) is another Hipparcos satellite 
photometric discovery. A new ephemeris, spectral type (G5) and V-light curve are  given by 
Gomez-Forrellad et al. (1999). The system has very shallow eclipses of equal depth, and 
unequal maxima (O'Connel effect). Duerbeck (1997) included this system in his list of 
contact binaries with low amplitude light variation. Observations of ET~Leo show 
that it undergoes marginal eclipses. Recently, Tanriverdi et al. (2004) obtained new 
photoelectric B and V light  curves and obtained a photometric solution of the system. 
It turned out to be a W-subtype contact binary with $q_{sp}$=0.342 $\pm$ 0.005 and a 
spectral type of G8V assigned by Rucinski et al. (2002).\\

{\it 2.6. ~XZ~Leo} \\
The eclipsing binary XZ~Leo (BD $+$17$^{\circ}$ 2165, HIP 49204, V = 10.29$^{m}$) was 
discovered by Hoffmeister (1934). Niarchos et al. (1994) gave the first times of minima, 
ephemeris and the physical parameters of the system. They noticed that the light curves 
were better approximated by introducing hot spots on both components near the neck region 
of the common envelope, to interpret the observed light curve asymmetries. The mass ratio, 
calculated by Rucinski and Lu (1999), is $q_{sp}$ = 0.348 $\pm$ 0.029, far away from 
the photometric one determined in previous studies. Its spectral type is A8/F0V and it 
belongs to the A-subtype of contact binaries.\\

{\it 2.7. ~V839~Oph} \\
The light variability of V839~Oph (BD $+$9$^{\circ}$ 3584, HIP 88946, V = 9.04$^{m}$) was 
discovered by Rigollet (1947). Several investigations followed, such as those made by 
Binnendijk (1960), Niarchos (1989), Akalin and Derman (1997).
Radial velocity 
and light curves analysis of the star was made by Pazhouhesh and Edalati (2003). 
They obtained new physical parameters of V839~Oph. Wolf et al. (1996) analyzed in detail 
the O-C diagram and tried to explain period changes in terms of conservative mass transfer 
or magnetic activity of the system or even a possible third body orbiting the binary.
According to Rucinski and Lu (1999) the system is an A-subtype contact binary with 
$q_{sp}$=0.305 $\pm$ 0.024 and a spectral type of F7V. \\

{\it 2.8. ~V2357~Oph} \\
V2357 Oph (GSC 979:339, HIP 82967, V = 10.51$^{m}$) was discovered by the Hipparcos mission. 
It was initially classified as a pulsating variable and the Hipparcos ephemeris refers to
the times of maxima. Rucinski et al. (2003) calculated an accurate mass ratio from their 
radial velocities observations. They found that $q_{sp}$ = 0.231 $\pm$ 0.010 and estimated 
its  spectral type as G5V. Assuming that the Hipparcos ephemeris is correct, 
Rucinski et al. (2003) classified the system as an A-subtype contact binary.\\

{\it 2.9. ~AQ~Psc} \\
The eclipsing binary AQ~Psc (BD $+$6$^{\circ}$ 203, HIP 6307, V = 8.66$^{m}$) was discovered 
by Sarma \& Radhakrishnan (1982). For this A-subtype contact binary very few observations 
exist, providing only times of minima and a good linear ephemeris of the system. Lu and 
Rucinski (1999) calculated the spectroscopic mass ratio of the system and found the 
value of $q_{sp}$ = 0.226 $\pm$ 0.002. They classified it as a contact binary of F8V 
spectral type.\\

{\it 2.10. ~VY~Sex} \\
The eclipsing binary VY~Sex (BD $-$1$^{\circ}$ 2452, HD 93917, V = 9.01$^{m}$) was recently 
discovered by Lasala-Garcia (2001) from a small, private observatory. It is one of the rare 
cases that, according to its brightness, it remained undetected for a long time. Rucinski 
et al. (2003) calculated the mass ratio of the system and found the value of 
$q_{sp}$ = 0.313 $\pm$ 0.005. They classified it as a W-subtype contact binary of F9.5V 
spectral type.\\

\begin{table}
\caption{Log of observations}
\begin{center}
\begin{tabular}{lcl}
\hline
System 	& Observatory 		& Dates                                            					\\
\hline
V410~Aur	& Univ. of Athens Obs.      	& 21, 22, 26, 29 Nov; 11, 12, 13, 20 Dec 2004; 4 Jan 2005   		\\
CK~Boo	& Univ. of Athens Obs.      	& 19, 20, 21, 23 Feb 2005						\\
FP~Boo	& Univ. of Athens Obs.	& 3, 5, 6, 9, 16, 17, 18 March; 15, 16 May 2005			\\
V921~Her	& Univ. of Athens Obs.      	& 11, 12, 13, 14, 25, 26, 27, 31 Aug; 5, 7, 8, 11, 13, 15 Sep 2004	\\
ET~Leo		& Univ. of Athens Obs.      	& 25 Mar 2003								\\
XZ~Leo		& Univ. of Athens Obs.      	& 29, 31 Jan; 15, 18, 29 Feb 2004					\\
V839~Oph	& Univ. of Athens Obs.      	& 26, 27, 28 May; 5, 9, 11 Jun; 6, 7, 8 Jul 2004			\\
V2357~Oph	& Kryoneri Astr. Station 	& 10, 11, 12 March 2005						\\
AQ~Psc		& Kryoneri Astr. Station	& 24 Nov; 20 Dec 2003						\\
VY~Sex	& Univ. of Athens Obs.	& 8, 9, 10, 11 May 2005						\\
\hline
\end{tabular}
\end{center}
\label{Date}
\end{table}

\begin{center}
{\bf 3. ~Photometric observations}
\end{center}

The observations of eight targets: V410~Aur, CK~Boo, FP~Boo, V921~Her, ET~Leo, XZ~Leo, 
V839~Oph and VY~Sex were obtained at the University of Athens Observatory, Athens, Greece.
The instruments used were the 0.40 m Cassegrain reflector and a ST-8 CCD camera. The CCD 
camera uses a Kodak KAF-1600 CCD chip, cooled by a two-stage Peltier element, 
which provides a CCD working temperature of $-30\,^{\circ}\mathrm{C}$ below ambient. 
The CCD chip has 1536$\times$1024 useful pixels of 9x9 microns, covering an area of 
15$\times$10 arcmin, or 22$\times$15 if a focal reducer is used. The CCD camera is 
equipped with a set of U, B, V, R and I Bessell filters. 

V2357~Oph and AQ~Psc were observed at the Kryoneri Astronomical Station of the 
National Observatory of Athens, Greece. The 1.22 m Cassegrain 
reflector and a Photometrics CH250 CCD camera were used. The CCD camera uses a Class I, 
SI502 CCD chip, cooled by a three-stage Peltier element 
and traditional chilled water cooling that provides a stable CCD working temperature 
of $-40\,^{\circ}\mathrm{C}$. The CCD chip has 512$\times$512 useful pixels of 
24$\times$24 microns, covering an area of 2.5$\times$2.5 arcmin. The CCD camera is 
equipped with a set U, B, V, R and I Bessell filters. The log of observations for 
all objects analyzed in this paper is given in Table 1.

One of the aims of this project is to obtain a complete light curve of each object at 
only one observatory to avoid the procedure of combining data taken at different 
sites and with different instruments. Therefore, each target has been observed with the 
same instrument and under the same configuration. In addition, the need to minimize 
intrinsic variations we attempted to complete the light curves in as short as possible time.

All data were left in the instrumental system. However, for the light curve modelling we 
transformed differential magnitudes into flux units. The observations were phased using 
linear ephemeris for all targets, taken from the Kreiner's (2004) catalogue, which is 
available on-line at: {\it http://www.as.ap.krakow.pl/ephem}.  Table 2 lists the reference 
epochs, as well as the periods used for phasing our new photometric observations.  \\

\begin{table}
\begin{center}
\caption{Linear elements used for phasing the observations}
\begin{tabular}{lcl}
\hline
System   	& reference epoch (HJD) & period (days)  		\\
\hline
V410~Aur	& 2452500.0033   		& 0.36635614 		\\
CK~Boo	& 2452500.0237		& 0.35515534		\\
FP~Boo	& 2452500.2796 		& 0.64048195		\\
V921~Her	& 2452500.0973     		& 0.87737804		\\
ET~Leo		& 2452500.3582      		& 0.34650349		\\
XZ~Leo		& 2452500.4192   		& 0.48773790 		\\
V839~Oph	& 2452500.4447     		& 0.40900491 		\\
V2357~Oph 	& 2452500.0907 		& 0.41556707		\\
AQ~Psc		& 2452500.3638		& 0.47561168 		\\
VY~Sex 	& 2452500.1065 		& 0.44343192		\\
\hline
\end{tabular}
\label{Efem}
\end{center}
\end{table}

\begin{center}
{\bf 4. ~Light curve modelling}
\end{center}

The method used for the derivation of the physical parameters is described in detail in 
Paper~I. The procedure described in Paper~II aims to the re-determination of the mass ratio 
for each system, correcting it for proximity effects. In order to obtain the physical 
parameters of components, we used the Wilson-Devinney code (W-D) (Wilson \& Devinney 1973; 
Wilson 1979, 1993) and applied the Monte Carlo algorithm as the search method.
Adjustments have been made to the following parameters: phase shift, inclination, 
temperature of the secondary component, potential(s) and the luminosity of the primary 
star. 

In our computations theoretical values of the  gravity darkening and albedo coefficients 
were used. In the cases where a star has a radiative envelope (T$>$7200\,K), we used the 
value of 1.0 for both coefficients, whereas in the case of convective envelopes (T$<$7200\,K) 
we set the albedo coefficient to 0.5 and the gravity darkening coefficient to 0.32. The 
limb darkening coefficients were adopted as functions of the temperature and wavelength 
from D\'{\i}az-Cordov\'es et al.~(1995) and Claret et al.~(1995) tables. The tables by 
Harmanec (1988) provided us with the effective temperature of the primary component, 
according to the spectral type obtained at DDO. Table 3 shows the searching ranges for 
all free parameters. In Tables 4 and 5 the results from the light curve modelling are
given for all systems analyzed in this paper. For models with spots the resulting spots 
parameters are also presented. 

In the cases of an obvious O'Connell effect, a spot was considered in our solution. In 
such cases, the whole surface of the brighter component was scanned for a possible spot 
location.  The comparison between the resulting theoretical light curves and the observations
is shown in Figs. 1-10. By combining the photometric solution with the spectroscopic results 
we calculated the absolute parameters of the two components, which are presented in Table 6. 
The listed errors correspond to the 90~percent confidence level.\\

\begin{table}
\begin{center}
\caption{Searching ranges of the adjusted parameters (W-D program input values)}
\begin{tabular}{lccccc}
\hline
System   	& $\phi$ 	& $i$(deg)	&  $T_{\rm 2}({\rm K})$  & $\Omega_{\rm 1}$=$\Omega_{\rm 2}$ &	$L_{1}$ 	\\
\hline
V410~Aur	& -0.01-0.01	& 65-90	& 4500-7500		& 1.90-4.50		& 2-12.5	\\
CK~Boo	& -0.01-0.01	& 35-90	& 4700-7700		& 1.90-4.50		& 1-12.5	\\
FP~Boo	& -0.01-0.01	& 35-80	& 5500-8500		& 1.90-4.50		& 1-12.5	\\
V921~Her	& -0.01-0.01	& 65-90	& 6000-9500		& 1.90-4.50		& 2-12.5	\\
ET~Leo		& -0.01-0.01	& 35-90	& 4000-7000		& 4.00-9.50		& 1-12.5	\\
XZ~Leo		& -0.01-0.01	& 65-90	& 5500-8000		& 1.90-4.50		& 2-12.5	\\
V839~Oph	& -0.01-0.01	& 65-90	& 4500-7500		& 1.90-4.50		& 2-12.5	\\
V2357~Oph	& -0.01-0.01	& 35-80	& 4000-7000		& 5.90-9.50		& 1-12.5	\\
AQ~Psc		& -0.01-0.01	& 65-90	& 4500-7500		& 1.90-4.50		& 2-12.5	\\
VY~Sex 	& -0.01-0.01	& 50-90	& 5000-7000		& 4.00-9.50		& 1-12.5	\\
\hline
\end{tabular}
\label{range}
\end{center}
\end{table}

\begin{table*}
\caption{Results derived from the light curve modelling} \begin{flushleft}
\begin{small}   
\begin{tabular}{lrrrrrrr}
\hline
parameter                           	&  V410~Aur         		&  CK~Boo         		& FP~Boo			& V921~Her              		& ET~Leo			\\	
\hline								
filling factor                      	&  72\%               		&  91\%               		& 38\%                		& 23\%                   		& 55\%               		\\	
phase shift                        	&  0.9960 $\pm$0.0006	&  0.9850 $\pm$0.0004	& 0.9990 $\pm$0.0003      	& 0.9992 $\pm$0.0002	& 0.0017 $\pm$0.0004  	\\	
$i$(deg)                        	&  80.6 $\pm$1.0   		&  63.7 $\pm$0.3   		& 68.8 $\pm$0.2        		& 68.1 $\pm$0.1      		& 36.6 $\pm$0.3       		\\	
$T_{\rm 1}({\rm K})$                	&  * 5890            		&  * 6150             		& * 6980               		& * 7700               		&  * 5500              		\\	
$T_{\rm 2}({\rm K})$                	&  5983 $\pm$22      		&  6163 $\pm$10		& 6456 $\pm$14        		& 7346 $\pm$20       		& 5112 $\pm$28      		\\	
$\Omega_{\rm 1} $ = $\Omega_{\rm 2}$&  2.004 $\pm$0.005&  1.915 $\pm$0.002		& 1.922 $\pm$0.012     	& 2.304 $\pm$0.002		& 6.178 $\pm$0.004 		\\	
q$_{corr}$(m$_{\rm 2}$/m$_{\rm 1})$   & * 0.137         	& * 0.106     			& * 0.096         			& * 0.244               		& * 2.924             		\\	
\hline	
$L_{1}~(B)$			& 8.423 $\pm$0.018		& 10.363 $\pm$0.015		& 11.386 $\pm$0.024		& 				& 3.922 $\pm$0.018		\\	
$L_{1}~(V)$			& 8.944 $\pm$0.015		& 10.366 $\pm$0.016		& 11.365 $\pm$0.021		& 9.555 $\pm$0.051		& 3.849 $\pm$0.015 		\\	
$L_{1}~(R)$			& 8.837 $\pm$0.013		& 10.368 $\pm$0.016		& 11.262 $\pm$0.022		& 9.569 $\pm$0.048		& 3.802 $\pm$0.013		\\	
$L_{1}~(I)$			& 8.882 $\pm$0.010		& 10.394 $\pm$0.014		& 11.207 $\pm$0.022		& 9.494 $\pm$0.047		& 3.696 $\pm$0.010		\\	
$L_{2}~(B)$			& ** 1.932			& ** 2.203			& ** 1.181			& 				& ** 8.644			\\	
$L_{2}~(V)$			& ** 2.064			& ** 2.200			& ** 1.202			& ** 3.011			& ** 8.717			\\	
$L_{2}~(R)$			& ** 1.995			& ** 2.198			& ** 1.304			& ** 2.997			& ** 8.764			\\	
$L_{2}~(I)$			& ** 1.963			& ** 2.172			& ** 1.359			& ** 3.072			& ** 8.870			\\	
$L_{3}~(B)$			& 0.176 $\pm$0.010		& * 0				& * 0				& 				& * 0				\\	
$L_{3}~(V)$			& 0.124 $\pm$0.010		& * 0				& * 0				& * 0				& * 0				\\	
$L_{3}~(R)$			& 0.138 $\pm$0.010		& * 0				& * 0				& * 0				& * 0				\\	
$L_{3}~(I)$			& 0.137 $\pm$0.010		& * 0				& * 0				& * 0				& * 0				\\	
\hline								
$r_{1}~^{pole}$			& 0.5309 $\pm$0.0012	& 0.5469 $\pm$0.0012	& 0.5442 $\pm$0.0002	& 0.4798 $\pm$0.0002	& 0.2964 $\pm$0.0012	\\	
$r_{1}~^{side}$			& 0.5928 $\pm$0.0020	& 0.6172 $\pm$0.0020	& 0.6115 $\pm$0.0003	& 0.5212 $\pm$0.0003	& 0.3130 $\pm$0.0015	\\	
$r_{1}~^{back}$		& 0.6179 $\pm$0.0024	& 0.6406 $\pm$0.0024	& 0.6311 $\pm$0.0004	& 0.5476 $\pm$0.0004	& 0.3722 $\pm$0.0034	\\	
$r_{2}~^{pole}$			& 0.2286 $\pm$0.0016	& 0.2157 $\pm$0.0016	& 0.1958 $\pm$0.0003	& 0.2551 $\pm$0.0003	& 0.4690 $\pm$0.0011	\\
$r_{2}~^{side}$			& 0.2408 $\pm$0.0019	& 0.2277 $\pm$0.0020	& 0.2048 $\pm$0.0003	& 0.2666 $\pm$0.0003	& 0.5094 $\pm$0.0016	\\	
$r_{2}~^{back}$		& 0.3043 $\pm$0.0020	& 0.3030 $\pm$0.0085	& 0.2469 $\pm$0.0007	& 0.3069 $\pm$0.0006	& 0.5450 $\pm$0.0021	\\	
\hline
spot parameters		&				&				&				&				&				\\
\hline
co-latitude (deg) 		& 97 $\pm$2			& 126	$\pm$2   		& ---    				& ---				& 74.4 $\pm$4 		\\
longitude (deg) 		& 306 $\pm$2			& 84 $\pm$2			& ---				& ---				& 41.5 $\pm$2	  		\\
radius (deg)  			& 12.2 $\pm$0.2		& 56.8	$\pm$0.5		& ---				& ---				& 11.5 $\pm$1 		\\
temp. factor			& 0.73 $\pm$0.05  & 0.98	$\pm$0.02 			& --- 				& ---				& 0.61 $\pm$0.01 		\\
\hline
\hline
\end{tabular}
\end{small}   
\end{flushleft}
\begin{small}   
$*$~-~assumed,~~~~$**$~-~computed,
~~~~$L_{1}, L_{2}$: W-D program input values -- the subscripts 1 and 2 refer to the star being eclipsed at primary and secondary minimum, respectively.\\
Spot parameters refer to the brighter component.\\
\end{small}   
\end{table*}

\begin{table*}
\caption[ ]{Results derived from the light curve modelling}
\begin{flushleft}
\begin{small}   
\begin{tabular}{lrrrrr}
\hline
parameter                    	& XZ~Leo            		& V839~Oph             		& V2357~Oph			&  AQ~Psc  			& VY~Sex			\\	
\hline								
filling factor           		& 19\%              		& 53\%              		& 23\%  			& 44\%               		& 22\%  			\\	
phase shift                  	& 0.0007 $\pm$0.0003	& 0.9996 $\pm$0.0004 	& 0.0016 $\pm$0.0004 	& 0.0013 $\pm$0.0002	& 0.0005 $\pm$0.0002	\\
$i$(deg)                  		& 78.3 $\pm$0.2      		& 82.9 $\pm$0.6      		& 48.0 $\pm$0.6 		& 68.9 $\pm$0.2		& 65.2 $\pm$0.2		\\	
$T_{\rm 1}({\rm K})$         	& * 7240             		& * 6250             		& * 5780   			& * 6100           		& * 5960  			\\	
$T_{\rm 2}({\rm K})$         	& 6946 $\pm$13      		& 6349 $\pm$25      		& 5640 $\pm$25		& 6124 $\pm$15      		& 5756 $\pm$15  		\\	
$\Omega_{\rm 1}$ = $\Omega_{\rm 2}$& 2.506 $\pm$0.005 & 2.357 $\pm$0.002   		& 8.223 $\pm$0.002 		& 2.244 $\pm$0.005		& 6.706 $\pm$0.005		\\	
q$_{\rm corr}$(m$_{\rm 2}$/m$_{\rm 1})$ & * 0.336            	& * 0.294            		& * 4.329    			& * 0.231   			& * 3.171 			\\	
\hline								
$L_{1}~(B)$			& 9.115 $\pm$0.015		& 8.582 $\pm$0.051		& 2.643 $\pm$0.051		& 9.210 $\pm$0.091		& 3.445 $\pm$0.091		\\	
$L_{1}~(V)$			& 9.014 $\pm$0.016		& 8.636 $\pm$0.051		& 2.627 $\pm$0.051		& 9.220 $\pm$0.078		& 3.399 $\pm$0.078		\\	
$L_{1}~(R)$			& 8.976 $\pm$0.016		& 8.658 $\pm$0.048		& 2.618 $\pm$0.048		& 9.265 $\pm$0.069		& 3.367 $\pm$0.069		\\	
$L_{1}~(I)$			& 8.941 $\pm$0.014		& 8.694 $\pm$0.047		& 2.615 $\pm$0.047		& 9.286 $\pm$0.070		& 3.331 $\pm$0.070		\\	
$L_{2}~(B)$			& ** 3.451			& ** 3.985			&** 9.924			& ** 3.356			& ** 9.122			\\	
$L_{2}~(V)$			& ** 3.552			& ** 3.931			&** 9.939			& ** 3.346			& ** 9.167			\\	
$L_{2}~(R)$			& ** 3.590			& ** 3.909			&** 9.949			& ** 3.302			& ** 9.199			\\	
$L_{2}~(I)$			& ** 3.625			& ** 3.873			&** 9.952			& ** 3.280			& ** 9.236			\\	
$L_{3}~(B)$			& * 0				& * 0				& * 0				& * 0				& * 0				\\	
$L_{3}~(V)$			& * 0				& * 0				& * 0				& * 0				& * 0				\\	
$L_{3}~(R)$			& * 0				& * 0				& * 0				& * 0				& * 0				\\	
$L_{3}~(I)$			& * 0				& * 0				& * 0				& * 0				& * 0				\\	
\hline								
$r_{1}~^{pole}$			& 0.4546 $\pm$0.0003	&0.4782 $\pm$0.0001	&0.2486 $\pm$0.0010		&0.4911 $\pm$0.0005		&0.2743 $\pm$0.0010		\\	
$r_{1}~^{side}$			& 0.4892 $\pm$0.0004	&0.5208 $\pm$0.0001	&0.2595 $\pm$0.0012		&0.5369 $\pm$0.0006		&0.2871 $\pm$0.0011		\\	
$r_{1}~^{back}$		& 0.5178 $\pm$0.0005	&0.5537 $\pm$0.0002	&0.2973 $\pm$0.0022		&0.5655 $\pm$0.0009		&0.3278$\pm$0.0020		\\	
$r_{2}~^{pole}$			& 0.2778 $\pm$0.0003	&0.2825 $\pm$0.0001	&0.4817 $\pm$0.0009		&0.2595 $\pm$0.0005		&0.4610$\pm$0.0009		\\	
$r_{2}~^{side}$			& 0.2906 $\pm$0.0004	&0.2977 $\pm$0.0001	&0.5234 $\pm$0.0013		&0.2725 $\pm$0.0006		&0.4974$\pm$0.0012		\\	
$r_{2}~^{back}$		& 0.3298 $\pm$0.0006	&0.3546 $\pm$0.0003	&0.5486 $\pm$0.0016		&0.3229 $\pm$0.0015		&0.5260$\pm$0.0015		\\	
\hline
spot parameters		&				&				&				&				&				\\
\hline
co-latitude (deg) 		& --- 				& --- 				& 50 $\pm$2			& ---    	 	  		& 94 $\pm$2			\\
longitude (deg) 		& ---				& --- 				& 59 $\pm$2			& ---    	 	  		& 281 $\pm$2			\\
radius (deg)  			& --- 				& --- 				& 10 $\pm$1			& ---    	 	 		& 19 $\pm$1			\\
temp. factor			& ---	 			& --- 				& 0.55 $\pm$0.01		& ---    	 	  		& 0.83 $\pm$0.01		\\
\hline
\hline
\end{tabular}
\end{small}   
\end{flushleft}
\begin{small}   
$*$~-~assumed,~~~~$**$~-~computed,
~~~~$L_{1}, L_{2}$: W-D program input values -- 
the subscripts 1 and 2 refer to the star being eclipsed at primary and secondary minimum, respectively. \\
Spot parameters refer to the brighter component.\\
\end{small}   
\label{TabRes}
\end{table*}

\begin{figure}
\begin{center}
\includegraphics[width=13cm,height=9.1cm,scale=1.0]{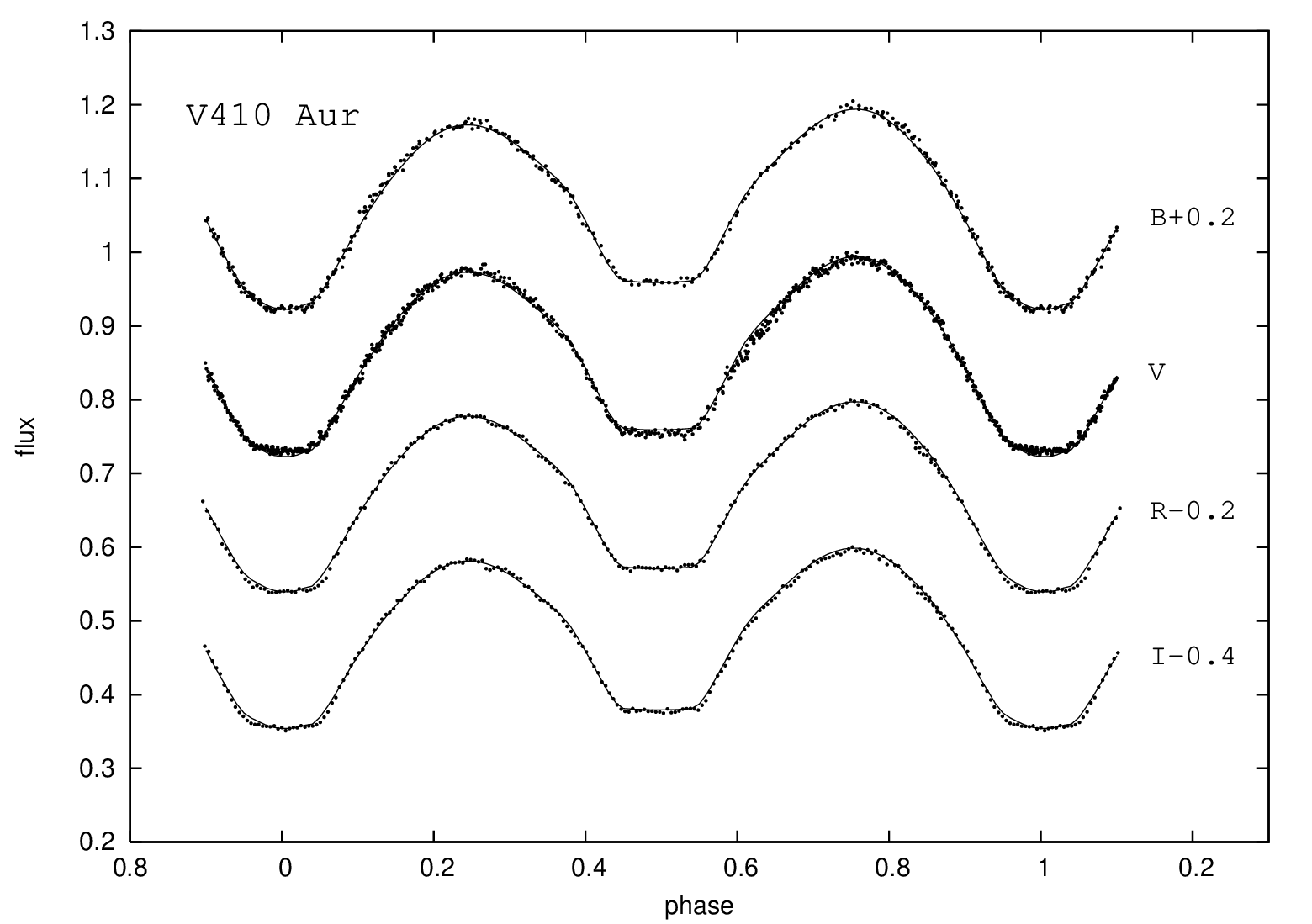}
\end{center}
\caption{Comparison between theoretical and observed light curves of V410~Aur 
(BVRI filters). Individual observations are shown with dots and theoretical curves with solid lines.}
\label{FigV417Aql}
\end{figure}

\begin{figure}
\begin{center}
\includegraphics*[width=13cm,height=9.1cm, scale=1.0,angle=0]{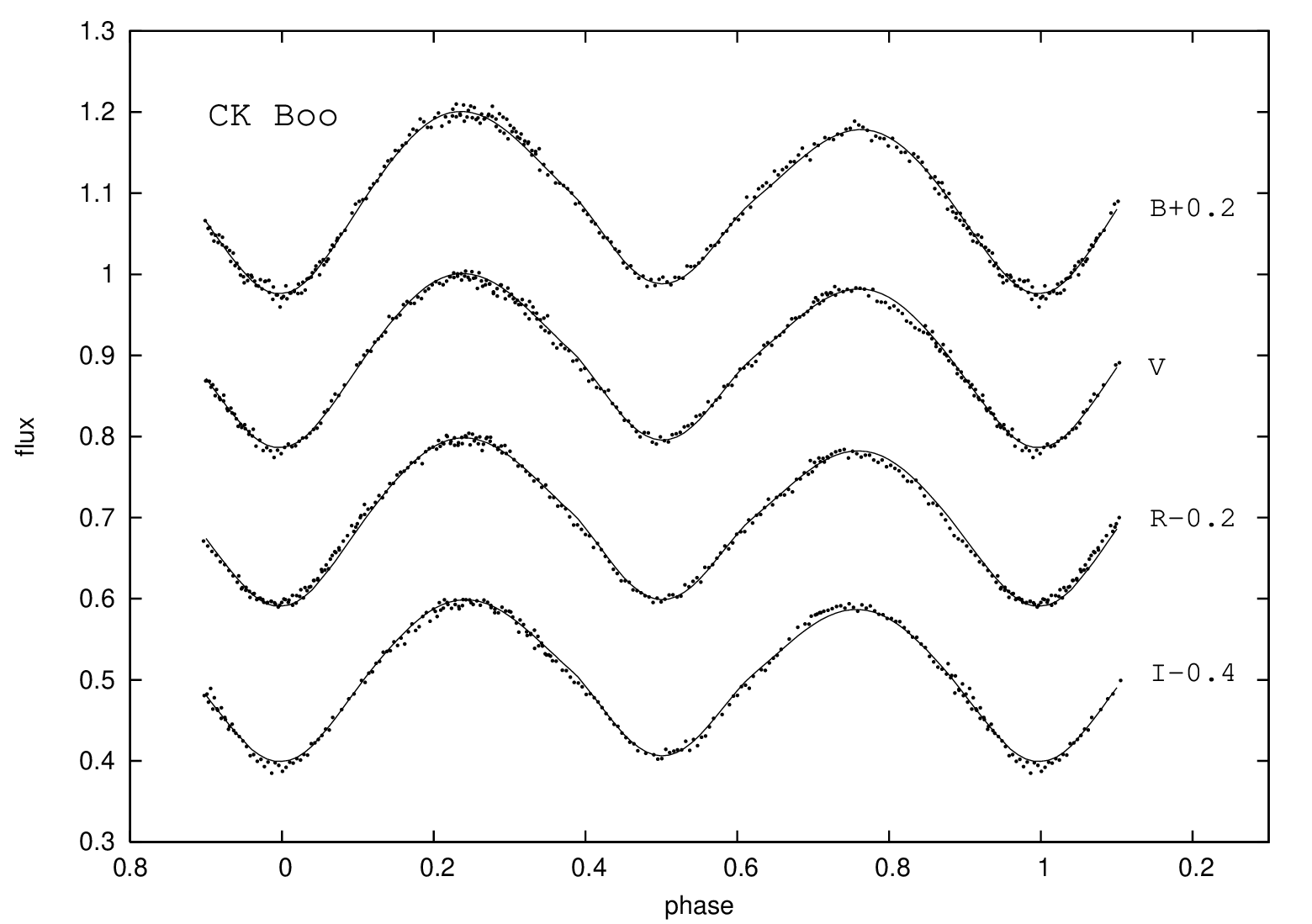}
\end{center}
\caption{Comparison between theoretical and observed light curves of CK~Boo 
(BVRI filters). Individual observations are shown with dots and theoretical curves with solid lines.}
\label{FigAHAur}
\end{figure}

\begin{figure}
\begin{center}
\includegraphics[width=13cm,height=9.1cm,scale=1.0]{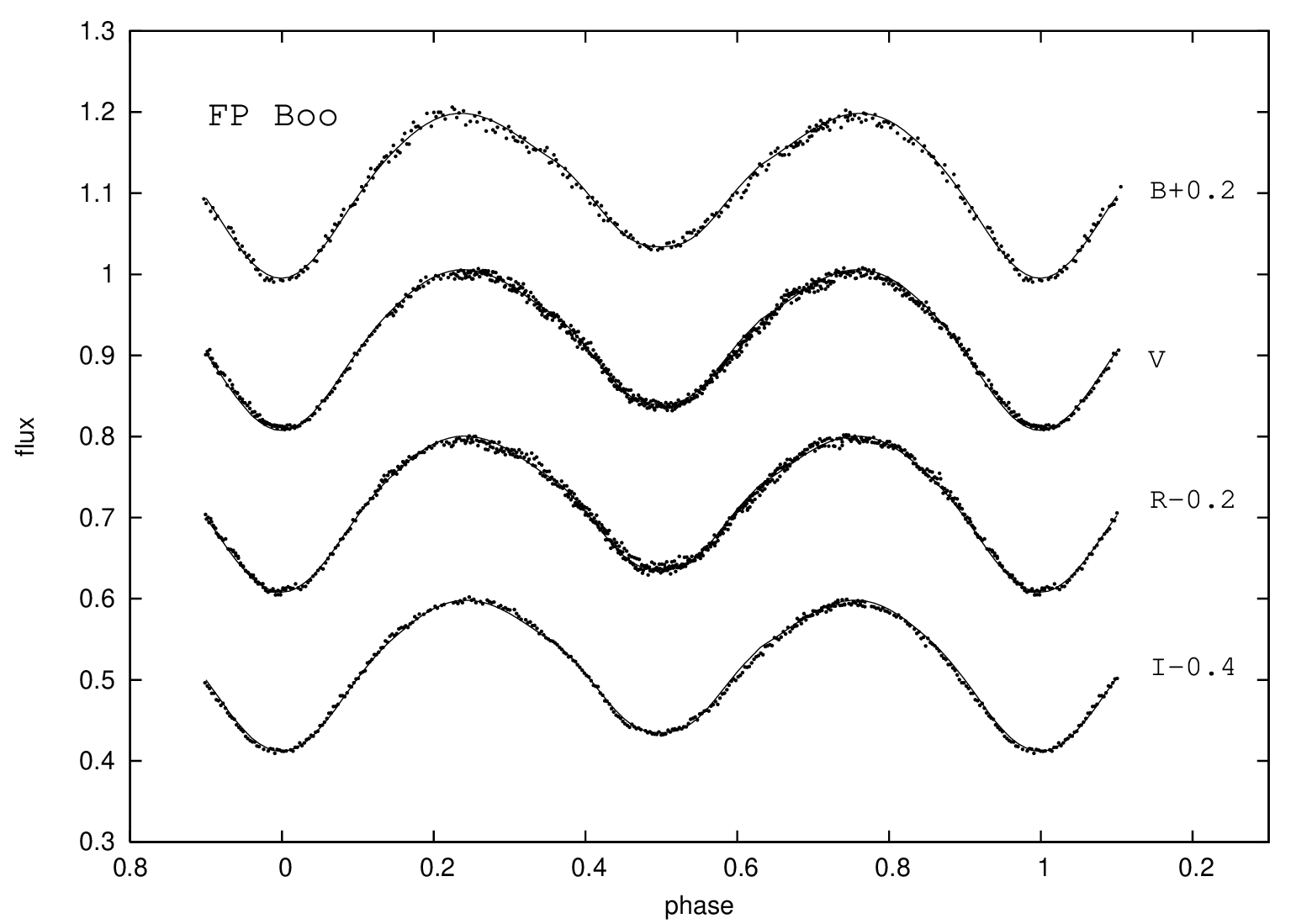}
\end{center}
\caption{Comparison between theoretical and observed light curves of FP~Boo 
(BVRI filters). Individual observations are shown with dots and theoretical curves with solid lines.}
\end{figure}

\begin{figure}
\begin{center}
\includegraphics[width=13cm,height=9.1cm,scale=1.0]{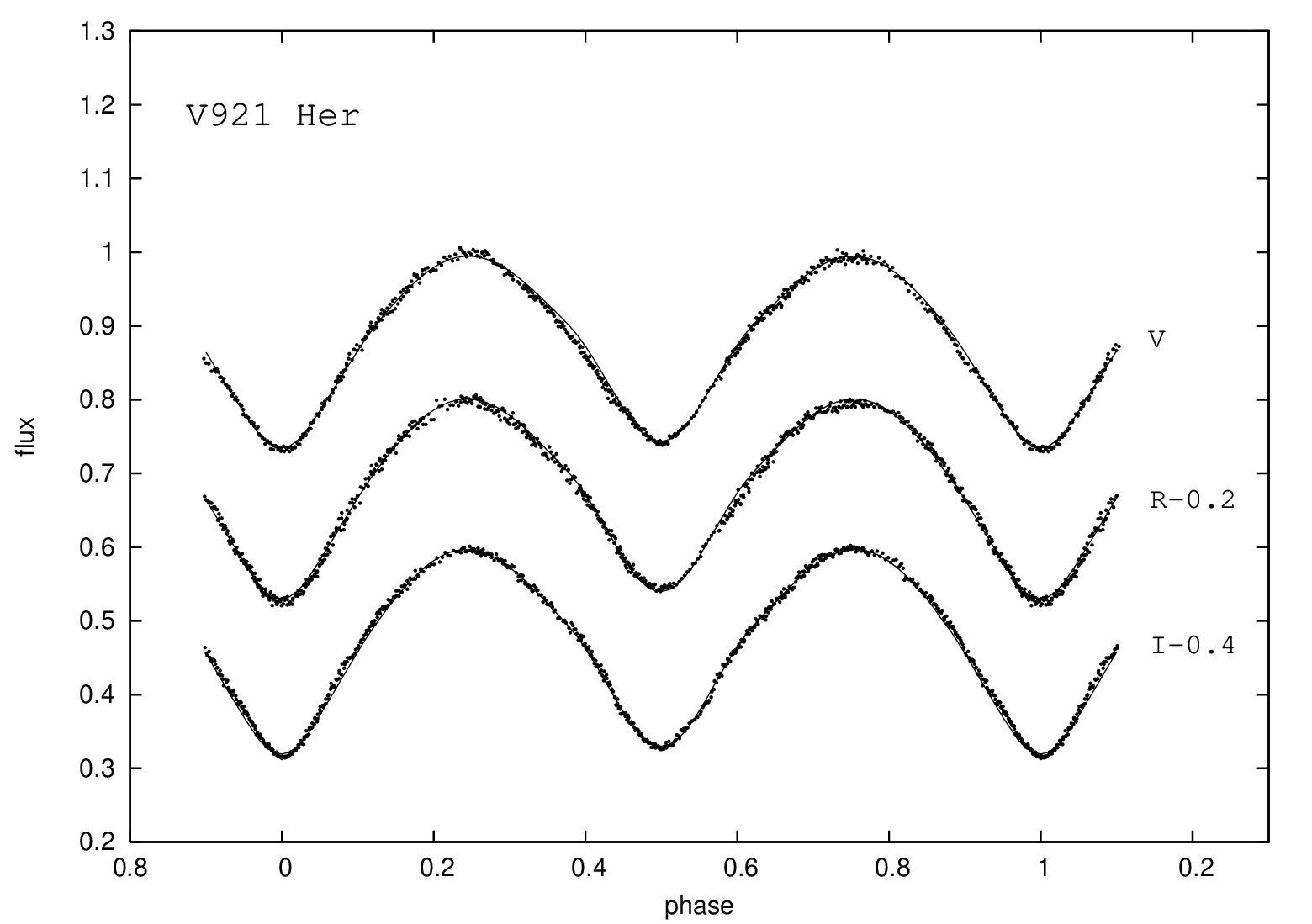}
\end{center}
\caption{Comparison between theoretical and observed light curves of V921~Her 
(VRI filters). Individual observations are shown with dots and theoretical curves with solid lines.}
\end{figure}

\begin{figure}
\begin{center}
\includegraphics[width=13cm,height=9.1cm,scale=1.0]{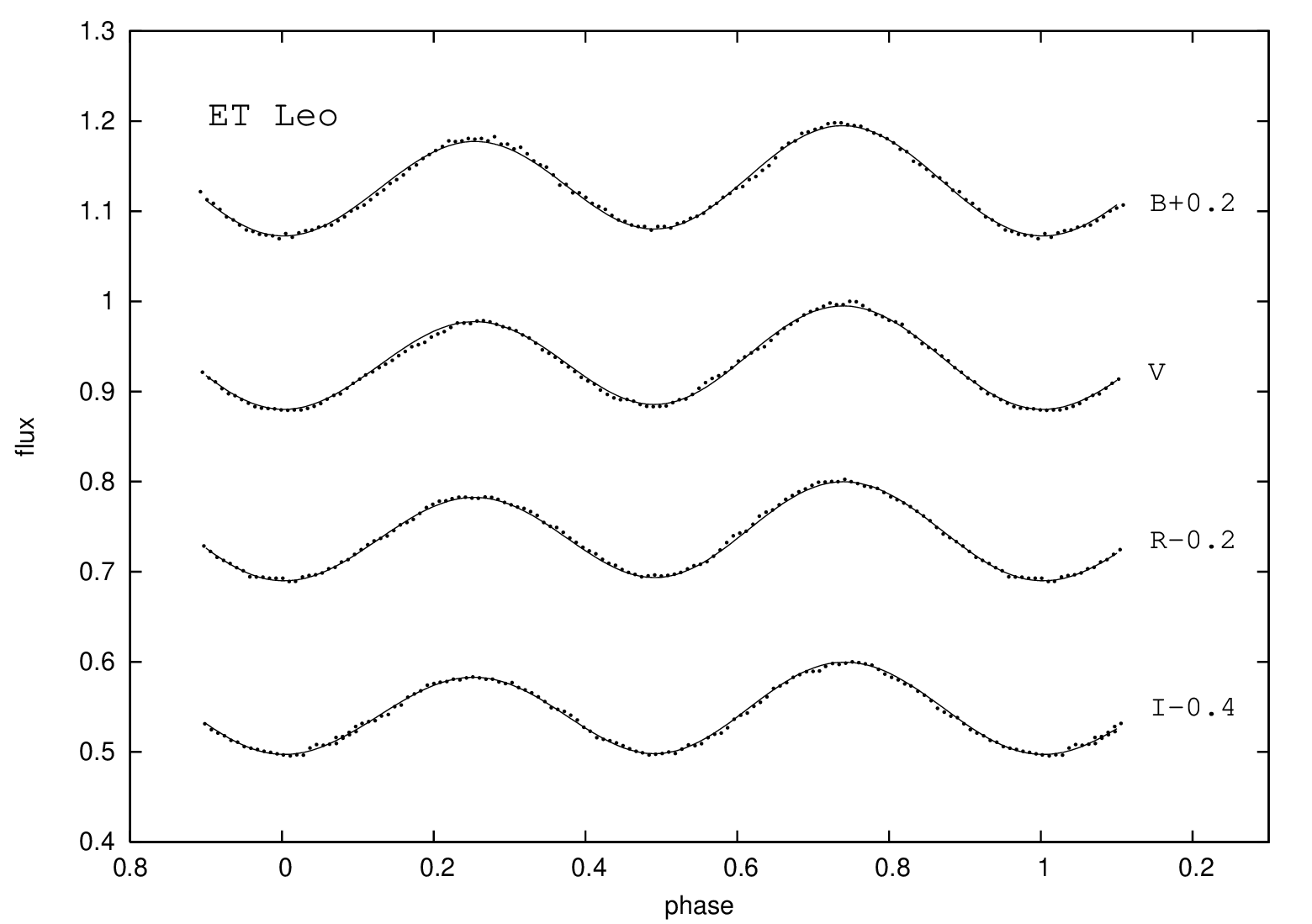}
\end{center}
\caption{Comparison between theoretical and observed light curves of ET~Leo 
(BVRI filters). Individual observations are shown with dots and theoretical curves with solid lines.}
\end{figure}

\begin{figure}
\begin{center}
\includegraphics[width=13cm,height=9.1cm,scale=1.0]{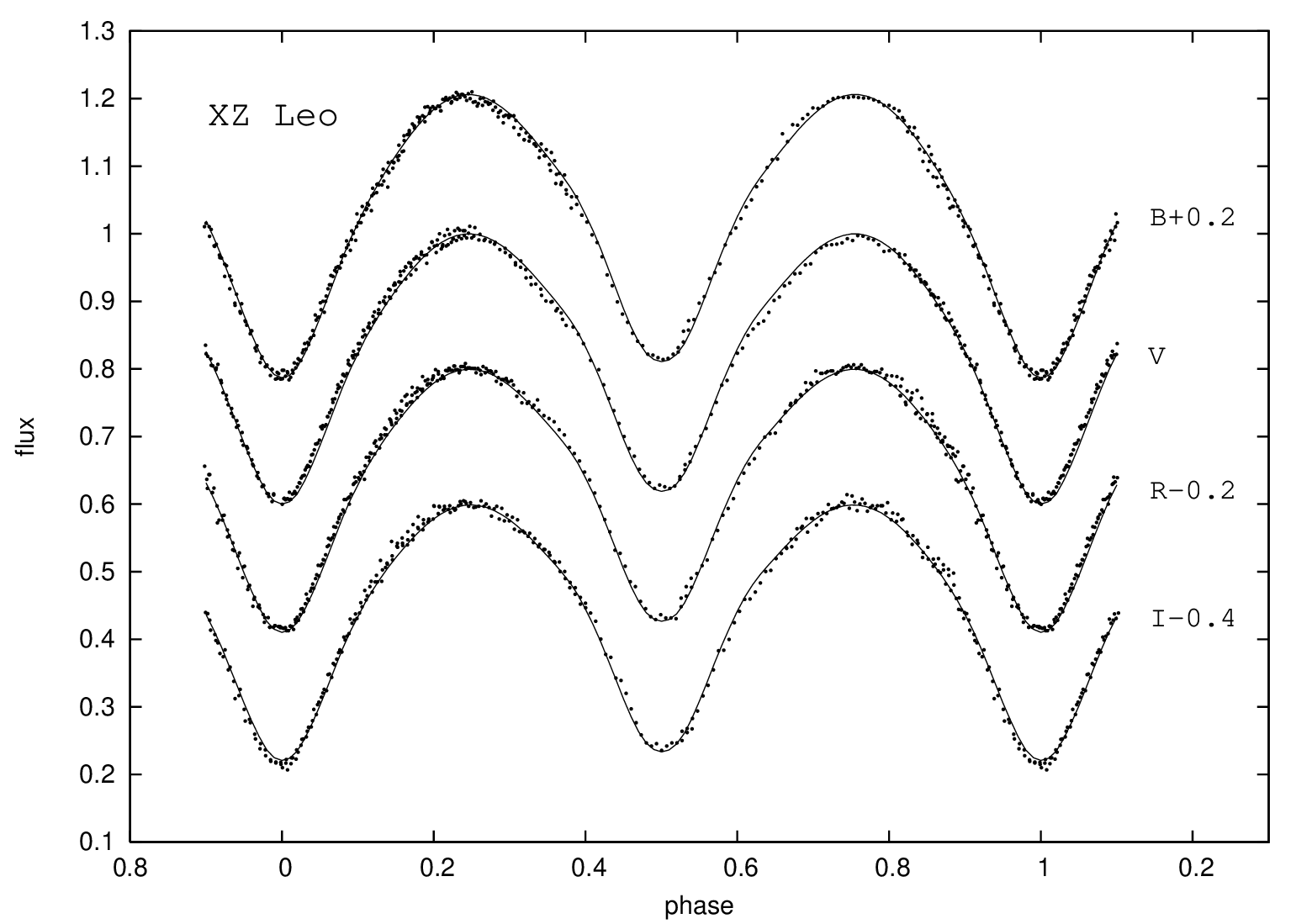}
\end{center}
\caption{Comparison between theoretical and observed light curves of XZ~Leo
(BVRI filters). Individual observations are shown with dots and theoretical curves with solid lines.}
\end{figure}

\begin{figure}
\begin{center}
\includegraphics[width=13cm,height=9.1cm,scale=1.0]{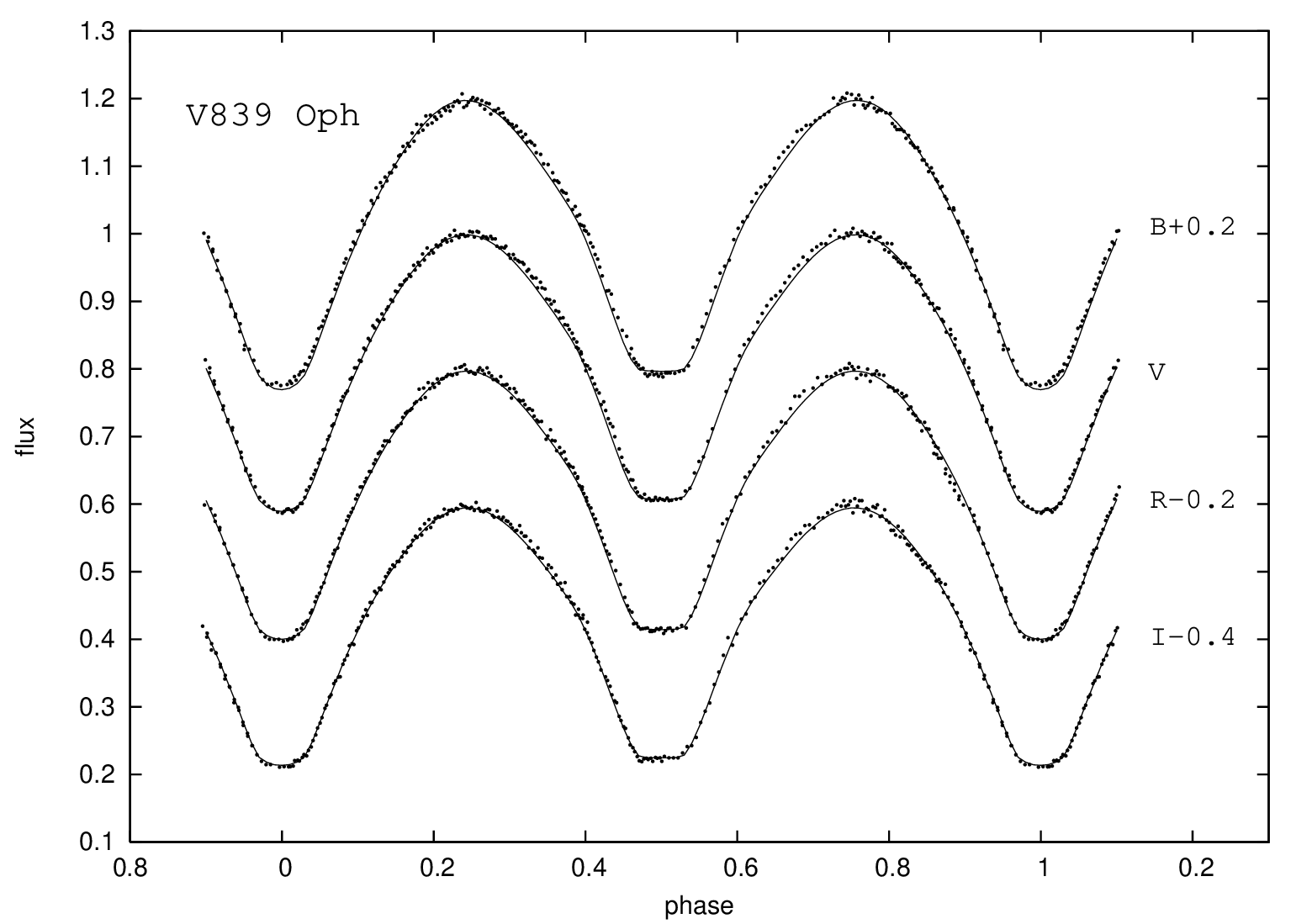}
\end{center}
\caption{Comparison between theoretical and observed light curves of V839~Oph 
(BVRI filters). Individual observations are shown with dots and theoretical curves with solid lines.}
\label{FigV839Oph}
\end{figure}

\begin{figure}
\begin{center}
\includegraphics[width=13cm,height=9.1cm,scale=1.0]{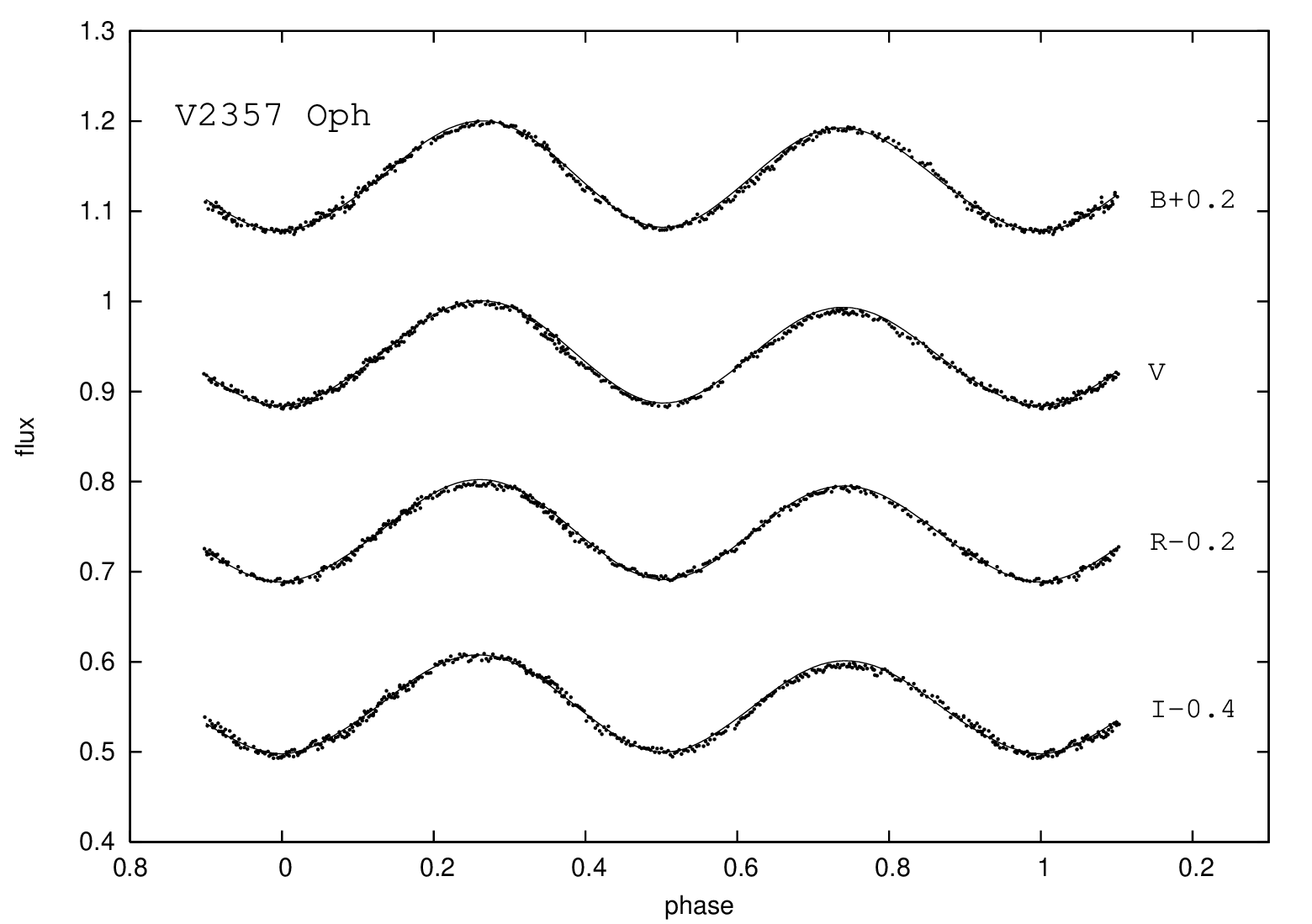}
\end{center}
\caption{Comparison between theoretical and observed light curves of V2357~Oph
(BVRI filters). Individual observations are shown with dots and theoretical curves with solid lines.}
\label{FigV2357Oph}
\end{figure}

\begin{figure}
\begin{center}
\includegraphics[width=13cm,height=9.1cm,scale=1.0]{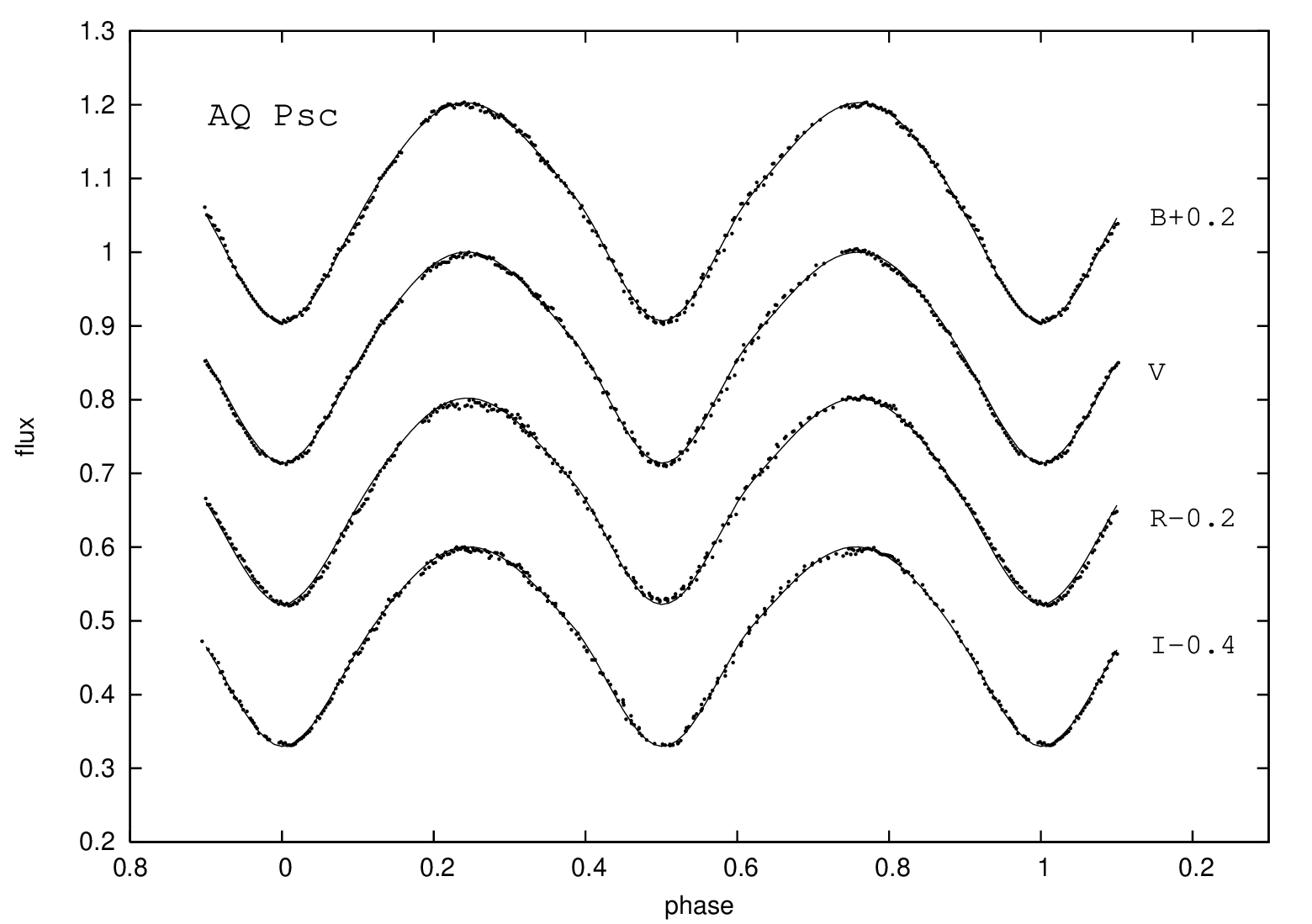}
\end{center}
\caption{Comparison between theoretical and observed light curves of AQ~Psc 
(BVRI filters). Individual observations are shown with dots and theoretical curves with solid lines.}
\label{FigAQPsc}
\end{figure}

\begin{figure}
\begin{center}
\includegraphics[width=13cm,height=9.1cm,scale=1.0]{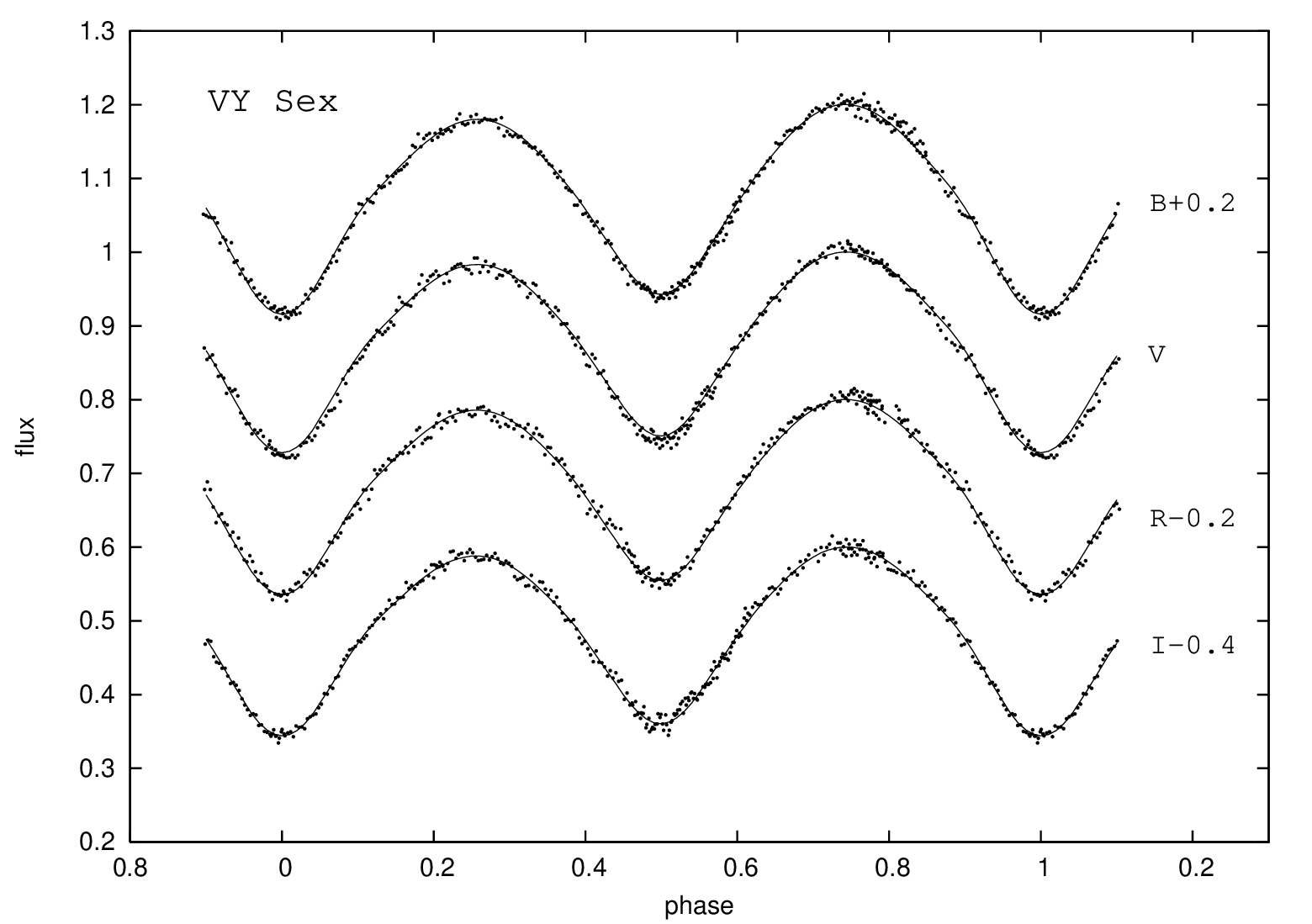}
\end{center}
\caption{Comparison between theoretical and observed light curves of VY~Sex
(BVRI filters). Individual observations are shown with dots and theoretical curves with solid lines.}
\label{FigVYSex}
\end{figure}

\begin{table}
\begin{center}
\caption[ ]{Absolute parameters (in solar units) of the systems studied in the present paper. The standard errors are expressed in parenthesis, in units of last decimal places quoted}
\begin{tabular}{lcccccc}
\hline
System    	& ${\cal M}_{\rm 1}$ 	& ${\cal M}_{\rm 2}$ 	& $R_{\rm 1}$ 		& $R_{\rm 2}$   	& $L_{\rm 1}$ 		& $L_{\rm 2}$ 		\\
\hline
V410~Aur	&	1.270(61)	&	0.173(25)	&	1.442(26)	&	0.586(11)	&	2.227(46)	&	0.389(15)	\\ 
CK~Boo	& 	1.442(24)	&	0.155(10)	&	1.521(10)	&	0.561(06)	&	2.924(20)	&	0.401(16)	\\ 
FP~Boo	& 	1.614(52)	&	0.154(21)	&	2.310(25)	&	0.774(08)	&    11.193(99)	&	0.920(13)	\\ 
V921~Her	& 	2.068(49)	&	0.505(26)	&	2.752(21)	&	1.407(11)	&    23.526(87)	&	5.094(35)	\\ 
ET~Leo	&	0.542(12)	& 	1.586(21)	&	0.835(07)	&	1.359(09)	&	0.564(08)	&	1.115(15)	\\ 
XZ~Leo	& 	1.742(47)	&	0.586(27)	&	1.689(15)	&	1.004(09)	&	6.926(43)	&	2.073(20)	\\ 
V839~Oph	& 	1.572(31)	&	0.462(17)	&	1.528(10)	&	0.874(06)	&	3.148(20)	&	1.097(10)	\\ 
V2357~Oph	&	0.288(09)	& 	1.191(12)	&	0.689(09)	&	1.392(18)	&	0.468(11)	&	1.782(30)	\\ 
AQ~Psc	& 	1.682(32)	&	0.389(17)	&	1.753(11)	&	0.890(05)	&	3.760(23)	&	0.984(08)	\\ 
VY~Sex	&	0.449(09)	& 	1.423(16)	&	0.864(06)	&	1.497(08)	&	0.832(08)	&	2.174(22)	\\ 
\hline
\end{tabular}
\end{center}
\end{table}

\newpage

\begin{center}
{\bf 5. ~Discussion and Conclusions}
\end{center}

We present the results of the combined photometric and spectroscopic solution for ten systems 
from the sample of close binary stars defined in Paper~I (Kreiner et al. 2003).
The solutions utilize new multicolor photometric data obtained through an international 
collaboration and results from homogeneous spectroscopic observations obtained within the 
David Dunlap Observatory Radial Velocity Program.  
We were able to derive the absolute physical parameters of  
components in contact systems with an accuracy of approximately $1\% - 2\%$. 
For half of the sytems analyzed in this paper spotted solutions are invoked to explain 
the observed asymmetries of their light curves.  

The configuration of all ten systems is contact. We found that the system V410~Aur 
has a high degree of contact with fillout factor of $72\%$.
V410~Aur was known to be a triple system from the time the spectroscopic observations were 
analyzed (Rucinski et al. 2003), where it was found that the third component 
contribution to the total light is about 26\%. The third light, added to the model in our 
photometric solution, was derived to be about 15\%. 
The difference between the two values of the third light contribution 
is significant. Spectroscopic detection of a third light can be a very sensitive tool in 
finding the presence of a third component. With special processing, 
third components as faint as 0.1\% of the total light can be detected 
that way (D'Angelo et al. 2006). However, when the third component is 
moderately bright, the ratio $L_{3}/(L_{1}+L_{2})$ will be usually corrupted by the 
very different spectral continuum level for the heavily rotationally 
broadened and blended spectrum of the contact binary and of the slowly 
rotating third star. In fact, what we observe for contact binaries is 
really not a continuum, but a pseudo-continuum, sometimes by some 10 - 
20\% lower than the totally unavailable real continuum. Thus, we should 
expect that spectroscopic determinations of $L_{3}/(L_{1}+L_{2})$ may normally be 
exaggerated and that photometric values should be preferred as totally 
free of this effect. The highly inclined orbit of V410~Aur implies total eclipses and a flat secondary minimum is obvious. 
The light curve has a small O'Connell effect, and a spotted solution 
is given. No other  peculiarities were found in this system.

CK~Boo has also a high degree of contact with fillout factor of $91\%$.
For this system we arrived at a higher fillout factor than the one derived by Kalci and Deman (2005).
The light curve of this target exhibits a large O'Connell effect and a spotted solution 
was assumed with a cool spot on the primary star. The low inclination orbit produces a smooth 
light curve, and the spotted solution fits our observational data very well.

The lowest inclination (below 50 degrees) was found for V2357~Oph and ET~Leo. 
The shape of the light curve of V2357~Oph due to inclination as low as 48 degrees shows
a low amplitude of light variation. It has a small O'Connell effect and our spotted 
solution was done assuming existence of a cool spot on the primary component. 
According to the assumption made by Rucinski et al. (2003) this 
system belongs to the A-subclass. Our photometric observations show that the minima 
referred as secondary are deeper contrary to the  assumption  made by Rucinski et al. (2003). 
By phasing the light curves according to the deeper minimum, the system turns out to be 
a W-subtype contact binary and thus the mass ratio used in the modelling was reversed. 
Our solution gave a  contact configuration with a fillout factor of $23\%$.
Even lower inclination (i=37 degrees) was found for ET~Leo. An O'Connell effect 
is visible as well. In our spotted solution, a relatively cool spot on the secondary 
component gave a  reasonable  fit to our observed light curves. Our solution gave a 
contact configuration with a fillout factor of $55\%$. However, for such a low orbital
inclination this solution is of lower significance than for other systems.

V921~Her is an eclipsing binary which contains large and very bright components. 
There is no evidence of any spot activity on the surface of the two components and the light curve is smooth with 
deep minima and partial eclipses. Both components have rather high temperatures and a 
radiative envelope was assumed in the model. Our solution gave a low contact configuration 
with a fillout factor of $23\%$.

V839~Oph has almost total eclipses, a fact resulting from its highly inclined orbit (82.9 
degrees). Continuous observations on 9 nights on May, June and July 2004 showed  that the 
system undergoes a continuous brightening, which is obvious from the increased value of 
differential magnitudes of the light curves. Light curve variations were also noticed 
in the past by other investigators (Wolf, 1996 and Akalin \& Derman, 1997), who gave an explanation based on the 
magnetic activity of the system. The brightening time rate gives information on the rearrangements of spots. 
What is important in such a case, is the overall brightening, even at eclipses, so that the whole system must get 
brighter. A similar case was also noticed by Rucinski \& Paczynski (2002) for a system in the Galactic Bulge region.
Our spotted solution resulted in a hot spot on the secondary component, 
close to the neck region, with temperature factor of 1.03, and a hot region on the primary, 
likely as a result of light reflected from the hot spot. The bright region around 
the neck can be regarded as an effect of mass/energy 
exchange between the components through the connecting neck of the common envelope 
(Van Hamme \& Wilson, 1985). Although our solution may not be unique, it explains very 
well the observations. In the present analysis, only the unspotted set of data 
(three nights on May 2004) was used to derive the absolute parameters. The whole set of data 
of this interesting target is the subject of a more detailed study of its spot activity 
(Gazeas et al. 2005b). Our solution gave a contact configuration with a fillout factor of $53\%$.

FP~Boo has no asymmetries in its light curve. Its relatively low amplitude is a result of 
the low inclination orbit. It has a very small mass ratio, large radial velocities 
and shows partial eclipses. The fillout factor derived from our solution is $38\%$.

The solution of the system VY~Sex gave a contact configuration with a fillout factor 
of $22\%$. 
Its light curve exhibits a small O'Connell effect and our spotted solution gave a cool spot 
on the equatorial zone of the primary component.

XZ~Leo  and AQ~Psc have no asymmetries in their light curves. They do not show any O'Connell 
effect and the eclipses are partial. We adopted solutions with unspotted surfaces and the derived 
results are quite satisfactory. The fillout factors found are $19\%$ and $44\%$ for 
XZ~Leo  and AQ~Psc, respectively.

The results for the physical parameters, derived in the present study, will be used 
together with those obtained in the previous papers of this series (Papers I-V), as 
well as with those expected from the next papers of the W UMa Program (Kreiner et al. 2003), 
for a global study of the W UMa-type systems in a broad context 
of contact binary evolution.\\

\begin{center}

\end{center}
\noindent

\noindent
{\bf Acknowledgements.} 
This project was supported by the Polish National Committee grant No.2 P03D 006 22 and by the 
Special Account for Research Grants No 70/3/7187 (HRAKLEITOS) and 70/3/7382 (PYTHAGORAS) of the
National and Kapodistrian University of Athens, Greece (for KG and PN). The project is co-financed within Op. 
Education by the ESF (European Social Fund) and National Resources. 
The research of SMR is supported by a grant from the Natural Sciences and Engineering 
Council of Canada. Support from SALT International Network grant No. 
76/E-60/SPB/MSN/P-03/DWM 35/2005-2007 is also acknowledged.\\

\begin{center}
REFERENCES
\end{center}
\noindent

\noindent
Akalin, A., and Derman, E. 1997, Astron. Astrophys. Suppl. Ser. {\bf 125}, 407.	 												 		\\																
Aslan, Z., and Derman, E. 1986, Astron. Astrophys. Suppl. Ser., {\bf 66}, 281.													 	 	\\																
Baran, A., Zola, S., Rucinski, S. M., Kreiner, J. M., Siwak, M., Drozdz, M., 2004, Acta Astron., {\bf 54}, 195 (Paper II). 										 	\\
Binnendijk, L. 1960, Astron. J., {\bf 65}, 79.						 												 		\\											
Bond, H.E. 1975, PASP, {\bf 87}, 877.							 												 		\\										
Claret, A., D\'{\i}az-Cordov\'es,  J., Gimenez, A., 1995,  Astron. Astrophys. Suppl. Ser., {\bf 114}, 247. 												\\	
D'Angelo, C., van Kerkwijk, M. H., Rucinski S. M., 2006, Astron. J., submitted (astro-ph/0602139)							\\
D\'{\i}az-Cordov\'es,  J., Claret, A., Gimenez, A., 1995,  Astron. Astrophys. Suppl. Ser., {\bf 110}, 329. 												\\
Duerbeck, H.W., 1997, IBVS 4513.							\\
Dvorak, S.W. 2005, IBVS, 5603.							 												 		\\										
ESA, 1997, The Hipparcos and Tycho Catalogues, ESA SP-1200, Noordwijk.	 												 		\\																
Gazeas, K. D., Baran, A., Niarchos, P., Zola, S., Kreiner, J. M., et al., 2005a, Acta Astron. {\bf 55}, 123 (Paper IV).		\\
Gazeas, K.D., Niarchos, P.G., Gradoula, G.-P., 2005b, Proceedings of the International Conference on Conatct Binaries, 27-30 June 2005, Syros, Greece (in press)				\\
Gomez-Forrellad, J.M., Garcia-Melendo, E., Guarro-Flo, J., Nomen-Torres, J., and Vidal-Sainz, J. 1999, IBVS, 4702.									 	\\														
Harmanec, P., 1988, Bull. Astron. Inst. Czechosl., {\bf 39}, 329 	\\			
Hoffmeister, C. 1934, Astron. Nachr., {\bf 253}, 195.			\\
Kalci, R., Derman, E., 2005, Astron. Nachr., {\bf 326}, 342.	\\
Kreiner, J. M., Rucinski, S. M., Zola, S., Niarchos, P., Ogloza, W., et al., 2003, Astron. Astrophys., {\bf 412}, 465 (Paper I). \\
Kreiner, J.M., 2004, Acta Astron., {\bf 54}, 207. 																			 	\\			
Krzesinski, J., Mikolajewski, M., Pajdosz, G., and Zola, S. 1991, Astrophys. Space Sci. {\bf 184}, 37.											 	\\												
Lasala-Garcia, A. 2001, IBVS, 5075.														 						 	\\			
Lu, W., and Rucinski, S.M. 1999, Astron. J., {\bf 118}, 515.											 						 	\\						
Niarchos, P.G. 1989, Astrophys. Space Sci., {\bf 153}, 143.											 						 	\\						
Niarchos, P.G., Hoffmann, M., and Duerbeck, H.W. 1994, Astron. Astrophys., {\bf 292}, 494.						 						 	\\											
Pazhouhesh, R., and Edalati, M.T. 2003, Astrophys. Space Sci., {\bf 288}, 259.								 						 	\\									
Pych, W., Rucinski, S. M., DeBond, H., Thomson, J. R., Capobianco, C. C., et al., 2004, Astron. J., {\bf 127}, 1712	\\
Qian, S., and Liu, Q. 2000, Astrophys. Space Sci., {\bf 271}, 331.										 						 	\\							
Rigollet, R. 1947, IAU Circ., 1013.														 						 	\\			
Rucinski, S.M., and Lu, W. 1999, Astron. J., {\bf 118}, 2451.											 						 	\\						
Rucinski, S.M., Lu, W., Capobianco, C.C., Mochnacki, S.W., Blake, R.M., et al., 2002, Astron. J., {\bf 124}, 1738.					\\												
Rucinski, S.M., Capobianco, C.C., Lu, W., DeBond, H., Thomson, J.R., et al., 2003, Astron. J., {\bf 125}, 3258.	\\																
Rucinski, S.M., Paczynski, B., 2002, IBVS 5321				\\
Rucinski, S.M., Pych, W., Ogloza, W., DeBond, H., Thomson, J.R., Mochnacki, S.W., et al., 2005, Astron. J., {\bf 130}, 767.					\\											
Sarma, M.B.K., and Radhakrishnan, K.R. 1982, IBVS 2073.																	 	\\
Selam, S.O. 2004, Astron. Astrophys., {\bf 416}, 1097																	 	\\
Tanriverdi, T., Senavci, H.V., Selam, S.O., and Albayrak, B. 2004, ASP Conf. Ser., {\bf 318}, 189.												 	\\					
van Hamme, W., Wilson, R.E. 1985, Astron. Astrophys., {\bf 125},  25. 	\\
Wilson, R.E., and Devinney, E.J. 1973, Astrophys. J., {\bf 182}, 539.																	 	\\	
Wilson, R.E. 1979, Astrophys. J., {\bf 234}, 1054. 																			 	\\
Wilson, R.E. 1993, Documentation of Eclipsing Binary Computer Model. 															 	\\	
Wolf, M., Satunova, L., and Molik, P., 1996, IBVS 4304.																	 	\\
Yang, Y.-G., Qian, S.-B., Zhu, L.-Y., 2005, Astron. J.,  {\bf 130}, 2252 																	\\
Zola, S., Rucinski, S. M., Baran, A., Ogloza, W., Pych, W., et al., 2004, Acta Astron., {\bf 54}, 299 (Paper III).  				\\
Zola, S., Kreiner, J. M., Zakrzewski, B., Kjurkchieva, D. P., Marchev, D. V., et al., 2005, Acta Astron., {\bf 55}, 389 (Paper V).		\\

\end{document}